\documentclass[twocolumn,superscriptaddress,showpacs,preprintnumbers,floatfix,aps]{revtex4-1}

\usepackage[pdftex]{color,graphicx}
\usepackage{latexsym,amsmath,amssymb,gensymb}
\usepackage{hyperref}

\begin{document}

\title{Evidence of a low temperature dynamical transition in concentrated microgels}

\author{Marco Zanatta}
\affiliation{Department of Computer Science, University of Verona, Strada le Grazie 15, 37134, Verona, Italy}

\author{Letizia Tavagnacco}
\affiliation{CNR-ISC and Department of Physics, Sapienza University of Rome, Piazzale A. Moro 2, 00185, Rome, Italy}

\author{Elena Buratti}
\affiliation{CNR-IPCF, Sede Secondaria di Pisa, Consiglio Nazionale delle Ricerche, Area della Ricerca, via G. Moruzzi 1, 56124 Pisa, Italy}

\author{Monica Bertoldo}
\email{monica.bertoldo@pi.ipcf.cnr.it}
\affiliation{CNR-IPCF, Sede Secondaria di Pisa, Consiglio Nazionale delle Ricerche, Area della Ricerca, via G. Moruzzi 1, 56124 Pisa, Italy}

\author{Francesca Natali}
\affiliation{CNR-IOM, Operative Group in Grenoble (OGG), c/o Institut Laue Langevin, 6 rue Jules Horowitz, BP 156, 38042 Grenoble cedex 9, France}

\author{Ester Chiessi}
\affiliation{Department of Chemical Sciences and Technologies, University of Rome Tor Vergata, Via della Ricerca Scientifica I, 00133 Rome, Italy}

\author{Andrea Orecchini}
\email{andrea.orecchini@unipg.it}
\affiliation{Department of Physics and Geology, University of Perugia and CNR-IOM, Via A. Pascoli, 06123, Perugia, Italy}

\author{Emanuela Zaccarelli}
\email{emanuela.zaccarelli@cnr.it}
\affiliation{CNR-ISC and Department of Physics, Sapienza University of Rome, Piazzale A. Moro 2, 00185, Rome, Italy}

\begin{abstract}
A low-temperature dynamical transition has been reported in several proteins. We provide the first observation of a ``protein-like'' dynamical transition in nonbiological aqueous environments. To this aim we exploit the popular colloidal system of poly-N-isopropylacrylamide (PNIPAM) microgels, extending their investigation to unprecedentedly high concentrations. Owing to the heterogeneous architecture of the microgels, water crystallization is avoided in concentrated samples, allowing us to monitor atomic dynamics at low temperatures. By elastic incoherent neutron scattering and molecular dynamics simulations we find that a dynamical transition occurs at a temperature $T_d\sim250$~K, independently from PNIPAM mass fraction. However, the transition is smeared out on approaching dry conditions. The quantitative agreement between experiments and simulations provides evidence that the transition occurs simultaneously for PNIPAM and water dynamics. The similarity of these results with hydrated protein powders suggests that the dynamical transition is a generic feature in complex macromolecular systems, independently from their biological function.
\end{abstract}

\date{\today}

\maketitle

\section{Introduction}
Hydrated protein powders are known to undergo a so-called dynamical transition, which is associated to a sudden increase with temperature of the protein atomic mean-squared displacements (MSD). This is related to the onset of anharmonic motions which allow the protein to explore conformational substates corresponding to the structural configurations of functional relevance \cite{fenimore2004confsub}. Thus, the transition is accompanied by the activation of the protein functionality, making it the subject of intensive research for its high relevance in the biological context. On the other hand, the dynamical transition is distinct from the protein glass transition the latter taking place at a lower temperature and involving other types of atomic motions \cite{capaccioli2012evidence}.
Since its first observation in myoglobin \cite{doster1989dynamical}, the dynamical transition has been reported for several proteins with different structures \cite{capaccioli2012evidence,doster2008rev} and in more complex biomolecules such as lipid bilayers \cite{peters2017membranes} and DNA strands \cite{cornicchi2007dna}. Although it appears to be closely related to biological functionality \cite{rasmussen1992ribo}, some experiments have shown that the dynamical transition also occurs in unstructured proteins \cite{he2008} and even in mixtures of unbound amino acids in aqueous environment \cite{schiro2011protein}. Conversely, in dry biomolecules the transition is suppressed, thus highlighting the prominent role of hydration water.

Recent studies have shown that the dynamical transition takes place simultaneously for both protein and water dynamics \cite{schiro2015translational,wood2008coincidence}, thus lifting residual doubts on the close dynamical coupling between hydration water and biomolecules \cite{frauenfelder2009unified,Lehnert1998coupling,cornicchi2009coupling}. Together, these evidences put forward the hypothesis that water plays a driving role in the dynamical transition and thus suggest that this phenomenon may be ubiquitous in the context of hydrated systems.
This also opens up the possibility of investigating non-biological systems (for example synthetic macromolecules) to shed light on the nature of the transition and to unveil the role of water. To this aim, it is particularly important to focus on complex macromolecular environments that could, on one hand, mimic the multifaceted potential energy landscape of proteins and, on the other hand, avoid water crystallization and allow the investigation of atomic dynamics at low temperatures. These conditions can be achieved by using the peculiar characteristics of microgel particles \cite{fernandez2011microgel}. Microgels are colloids made by cross-linked polymer networks with a heterogeneous structure composed of a dense core and a loose corona.

\begin{figure*}[htb]
\centering
\includegraphics[width=0.95\textwidth]{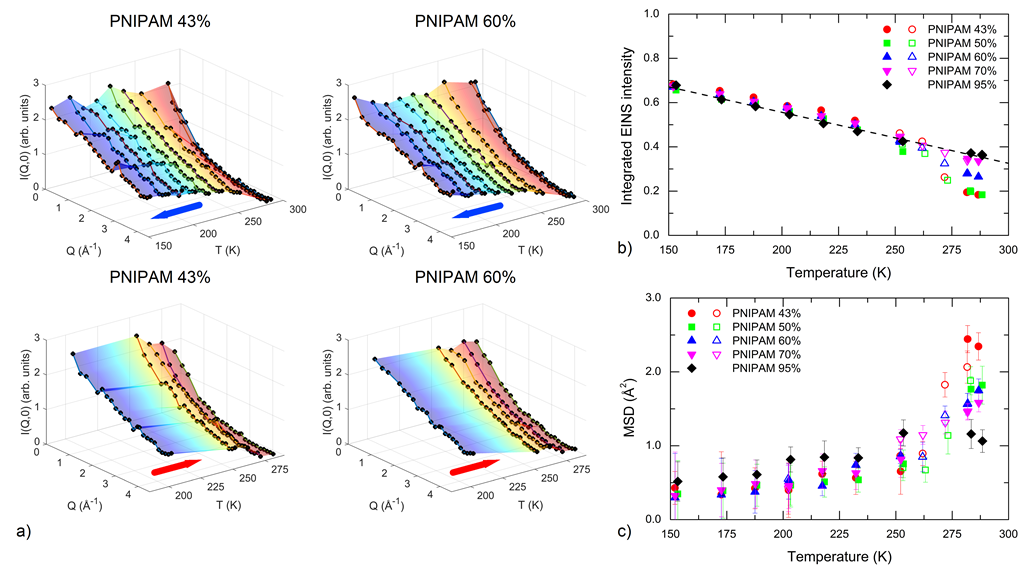}
\caption{\textbf{EINS data.} (a) Incoherent elastic intensities $I(Q,0)$ measured on PNIPAM microgels in D$_2$O with a mass fraction concentration of 43\% and 60\% as a function of temperature $T$. Top (bottom) panels reports data recorded under cooling (heating). (b) Integral over $Q$ of $I(Q,0)$ as a function of $T$. The integrated EINS intensities are normalized to 1 for $T\rightarrow 0$. (c) Temperature evolution of MSD as obtained using the double-well model (see SM). In (\textbf{B}) and (\textbf{C}) measurements under cooling are presented with filled symbols whereas those under heating with open symbols.}
\label{fig:EINS}
\end{figure*}

Poly-$N$-isopropylarcylamide (PNIPAM) microgels have so far been extensively studied around ambient temperature as a tunable model system for elucidating phase transitions and glassy behavior \cite{yunker2014physics}. We now extend their investigation to a yet unexplored region of phase diagram, encompassing PNIPAM mass fractions (wt) in the range $43\leq c\leq 95$ wt \% and a wide range of temperatures, namely $150\lesssim T\lesssim 290$~K.
By combining elastic incoherent neutron scattering (EINS) experiments with all-atom molecular dynamics (MD) simulations, we probe the internal dynamics of microgels at short time and length scales. We find that our suspensions do not crystallize for PNIPAM mass fractions $\gtrsim 43$ wt \% at any temperature, in good agreement with calorimetric data on PNIPAM chains \cite{afroze2000phase,van2004kinetics}.
Most crucially, we find the occurrence of  a ``dynamical'' transition at $T_d\approx 250$~K, akin to that observed in proteins. The value of the transition temperature does not depend on PNIPAM mass fraction, but the transition tends to disappear when approaching dry conditions. We directly compare the measured EINS intensities and their associated MSD with those calculated in simulations, finding quantitative agreement between the two. Moreover, simulations show a strong coupling for both PNIPAM and water dynamics, for which a discontinuity in diffusional dynamics occurs at the same temperature.
These results represent, to our knowledge, the first observation of a genuine dynamical transition in a nonbiological aqueous system, providing generality for the concept of a low-temperature dynamical transition in disordered macromolecules with internal degrees of freedom.

\section{Results}

\subsection{EINS experiments}
EINS experiments were performed on  PNIPAM microgels hydrated with D$_2$O at five polymer mass fractions between 43 and 95\% \cite{DataDOI}. In thermal neutron scattering, the incoherent cross section of hydrogen atoms is more than an order of magnitude larger than both coherent and incoherent cross sections of the other atomic species in our PNIPAM suspensions. Therefore, the incoherent signal of the hydrogen atoms in the PNIPAM network dominates the mostly coherent signal of deuterated water, providing selective access to the microscopic dynamics of the polymer matrix.

EINS data were collected at the backscattering spectrometer IN13 of the Institut Laue-Langevin (ILL; Grenoble, France). IN13 has an energy-resolution $\Delta E=8~\mu$eV (full width at half maximum) and covers a momentum transfer interval from 0.3 to 4.5~\AA$^{-1}$, thus accessing motions faster than about 150~ps taking place in a spatial region between 1 and 20~\AA. In this way, we essentially probe the internal dynamics of the microgel and its behavior on the atomic scale. This is important because, although our samples are glassy on the colloidal scale \cite{paloli2013fluid,bischofberger2015new}, the volume fraction occupied by the microgels is still much lower than that of the polymer glass transition \cite{afroze2000phase}; thus we can safely assume that the polymeric degrees of freedom are in equilibrium. Our results confirm this assumption, as they show a progressive and smooth behavior with PNIPAM mass fraction, are reproducible with respect to sample preparation and are fully reversible in temperature.

The measured incoherent elastic intensities $I(Q,0)$ as a function of temperature and momentum transfer $Q$ are shown in Fig.~\ref{fig:EINS}(a) for selected samples with PNIPAM mass fractions of 43 and 60\%. Data were recorded under thermal cycles of cooling (top panels) and heating (bottom panels) in a range of temperatures between 287 and 152~K. The temperature behavior is best observed in the $Q$ integral of $I(Q,0)$, as shown in Fig.~\ref{fig:EINS}(b). Upon cooling, the integrated EINS intensity progressively increases until a sharp variation occurs at a characteristic ``dynamical transition'' temperature $T_d \sim$250~K, below which the intensity increase continues with a smaller but linear slope. Such a discontinuity is more pronounced at the lowest PNIPAM concentration and gradually disappears with decreasing water content. Upon heating back, the measured data fall on top of the corresponding ones at the same $T$, confirming the full reversibility of the process.

At each temperature, the average atomic MSD can be calculated from the $Q$ dependence of $I(Q,0)$. To this end, the EINS data $I(Q,0)$ were fitted with a double-well model, which captures the data behavior very well (see figs.~\ref{fig:Iq40} and \ref{fig:Iq60}). The resulting MSDs in Fig.~\ref{fig:EINS}(c) reflect the discontinuity observed in the $Q$-integrated intensity at $T_d \sim$250~K. Below $T_d$, the linear temperature evolution of the MSD is typical of a harmonic solid. Conversely, the sudden slope increase above $T_d$ witnesses an enhanced mobility of the PNIPAM atoms, due to the onset of anharmonic motions and resulting in the corresponding elastic intensity drop.

It is worth noting that the observed discontinuity at $T_d$ might be attributed to an underlying crystallization process associated to the D$_2$O. However, the measured $I(Q,0)$ for pure D$_2$O in the same temperature range shows defined crystalline peaks that do not appear in low-$T$ PNIPAM suspensions (see Supplementary Materials, fig.~\ref{fig:D2O}). This suggests that the elastic intensity drop of the PNIPAM network originates from a mechanism other than water crystallization.

\subsection{MD Simulations}
\begin{figure}[htb]
\centering
\includegraphics[width=0.90\linewidth]{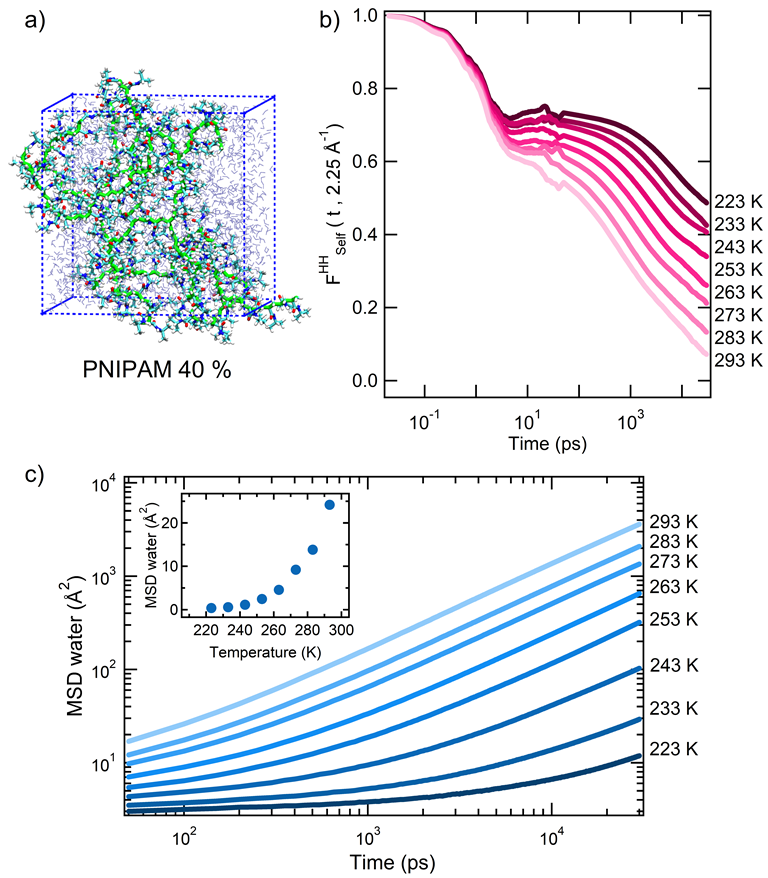}
\caption{\textbf{MD simulation  results for 40\% mass fraction.} (\textbf{A}) Snapshot of the simulated PNIPAM microgel. Chain atoms are displayed in green to highlight the network structure. (\textbf{B}) Self-intermediate scattering function of PNIPAM hydrogen atoms calculated at $Q=2.25$~\AA$^{-1}$ as a function of temperature. (\textbf{C}) Time-evolution of the mean square displacements of water molecules.
The inset shows the water mean square displacements values at 150~ps as a function of $T$.}
\label{fig:pnipam}
\end{figure}

To identify the microscopic mechanism responsible for the observed change in PNIPAM dynamics taking place at $T_d$, we rely on all-atom MD simulations. To this aim, we designed a new microgel model (described in Materials and Methods and in the Supplementary Materials) and performed its numerical investigation at mass fractions of 40 and 60\%. A snapshot of the in silico microgel is displayed in Fig.~\ref{fig:pnipam}(a). The PNIPAM network was built with a cross-linker/monomer ratio of 1:28 which, taking into account the inhomogeneous density of a PNIPAM particle, describes an inner region (for example, the core region) of the microgel. The number of water molecules per monomeric unit is smaller than the experimentally determined value of hydration molecules for PNIPAM microgels~\cite{Ono2006} for both concentrations. Thus, we can safely assume that all water molecules in the simulations are hydration water and not bulk water.

\begin{figure}[htb]
\centering
\includegraphics[width=0.90\linewidth]{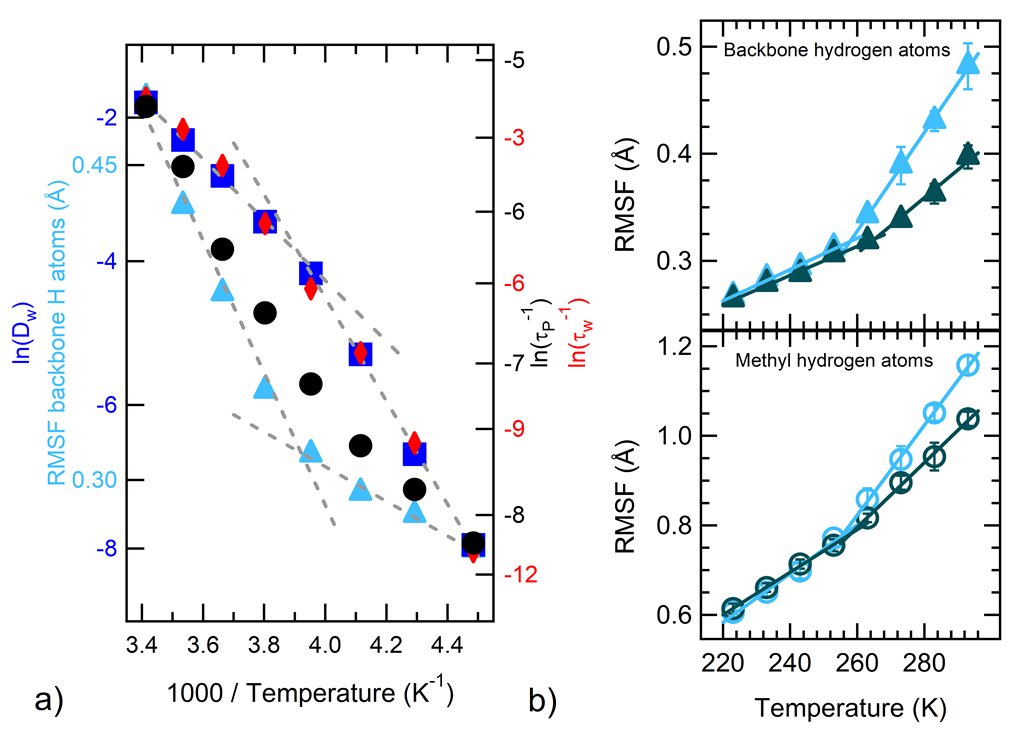}
\caption{\textbf{Water-PNIPAM coupling.} (a) Temperature dependence of water diffusion coefficient (squares); SISF relaxation times for PNIPAM hydrogen atoms (circles) and water oxygen atoms (diamonds); root mean square fluctuation (RMSF) of backbone hydrogen atoms (triangles) calculated from MD simulations of PNIPAM 40\% wt. Dashed lines are guides to the eye. Note that data sets are represented on different scales on the y-axis to improve visualization. (b) Temperature evolution of RMSF of the PNIPAM hydrogen atoms for 40\% wt (light blue) and for 60\% wt (dark green) averaged over 150ps: methyl groups (circles), backbone atoms (triangles).}
\label{fig:water}
\end{figure}

At each studied temperature, we monitored the dynamics of both microgel and water atoms separately. To characterize microgel dynamics we calculated the self intermediate scattering function (SISF) that probes the single-particle translational dynamics at a characteristic wave vector $Q$. Figure~\ref{fig:pnipam}(b) displays the SISFs calculated for the hydrogen atoms of PNIPAM 40 wt \% at $Q = 2.25$~\AA$^{-1}$, that is, at the position of the first peak in the oxygen-oxygen structure factor of bulk water\cite{sciortino1996supercooled}. PNIPAM internal dynamics exhibits a two-step behavior typical of glass-forming liquids: An initial fast relaxation is followed by a long-time, slow relaxation indicating structural rearrangement.  At the lowest studied temperatures, the SISFs do not decay completely to zero, an indication that the system is becoming arrested on the considered time window. However, studies on longer time scales reveal that aging phenomena do not play a major role at the studied temperatures (see figs.~\ref{fig:msd_water} and \ref{fig:msdPNIPAM}). The long-time relaxation of the SISFs as a function of time $t$ is well described by a stretched exponential as in standard glass formers, thus providing an estimate of the structural relaxation time $\tau_p$.  This is shown in Fig.~\ref{fig:water}(a) as a function of temperature in an Arrhenius plot. We find that $\tau_p$ obeys two distinct dynamical regimes, each being compatible with an Arrhenius dependence. The crossover temperature of about 250~K is strikingly similar to that found in EINS experiments. The activation energies are $28.4\pm0.8$~kJ~mol$^{-1}$ and $14.5\pm0.5$~kJ~mol$^{-1}$ for the  high and low temperature regime, respectively.

To complement these results, we also investigated the hydration water dynamics by looking at the MSD of the oxygen atoms, see Fig.~\ref{fig:pnipam}(c). Similar to PNIPAM SISFs, we observe a slowing down of  water dynamics with decreasing $T$. While at high $T$ the MSD shows a diffusive behavior in the studied time interval, upon lowering temperature the onset of an intermediate plateau is observed. This is a characteristic feature of glassy systems, indicating that atoms remain trapped in cages of nearest neighbors for a transient, before eventually diffusing away at long times. Water oxygen atoms retain a diffusive behavior at all studied $T$, so that it is possible to estimate their self-diffusion coefficient $D_w$. This is also shown in the Arrhenius plot of Fig.~\ref{fig:water}(a), where $D_w$ is also found to follow two distinct regimes, each one compatible with an Arrhenius dependence, crossing again at a temperature $\approx250$~K.
Furthermore, in the inset of Fig.~\ref{fig:pnipam}(c) the water MSD calculated at a time of 150~ps, matching the experimental time resolution, is shown as a function of $T$. A clear increase of the MSD is found above $\sim250$~K. This strikingly matches the MSD behavior observed by EINS in our PNIPAM samples and is closely reminiscent of other experimental and simulation results on protein hydration water~\cite{schiro2015translational,wood2008coincidence,combet2012pccp,chen2006observation,doster2010dynamical}.

To make a closer connection to protein dynamics, we calculated the root mean square fluctuation (RMSF) of PNIPAM hydrogen atoms, which measures the fluctuations of the atom positions with respect to the averaged structure over a defined period of time.
We monitored this quantity separately for hydrogen atoms of the methyl groups and those of the backbone. The evolution with temperature of the RMSF for both types of hydrogen atoms, averaged over the experimental resolution time, is shown in Fig.~\ref{fig:water}(b). Again, a clear change is detected at a temperature of about 250~K for both types of atoms with a stronger effect for the backbone hydrogen atoms. Such a change is observed for both studied mass fractions, suggesting that concentration does not play any role on the value of the crossover temperature, for both water and PNIPAM atoms, but it only smears out the variation, making the two regimes progressively similar to each other. These findings are in agreement with the concentration dependence of the experimental $I(Q,0)$ (Fig.~\ref{fig:EINS}(b)). Figure~\ref{fig:water}(a)  summarizes the MD results, showing that  $\tau_p$, the RMSF of PNIPAM backbone atoms, the self-diffusion constant of water and the associated relaxation time $\tau_w$ all display a clear change at  $T\sim250$~K, a temperature strikingly similar to the one where the EINS data reveal a variation. Thus, simulations show that the dynamical properties of both PNIPAM and water are slaved to each other.

These results suggest to interpret the transition observed at $T_d$ as the analog of the protein dynamical transition for PNIPAM microgel suspensions. Several features are shared by protein and PNIPAM dynamical transition: (i) $T_d$ does not depend on the concentration and the transition vanishes for dry conditions; (ii) the transition takes place at a temperature higher than that of the glass transition~\cite{capaccioli2012evidence}; (iii) the strong coupling between water and PNIPAM dynamics, as shown in Fig.~\ref{fig:water}(a), is well established in hydrated proteins~\cite{frauenfelder2009unified,Lehnert1998coupling,cornicchi2009coupling,schiro2015translational}.

It is then natural to ask whether water, through hydrogen bonding interactions with the involved macromolecule, drives the occurrence of this dynamical transition, that appears ubiquitous in proteins and in PNIPAM microgels. To partially answer this question, we refer to recent simulation studies on confined water, which have provided a detailed characterization of slow dynamics in water~\cite{gallo2010dynamic}. In particular the study of hydrated lysozyme~\cite{camisasca2016two} has identified a new (long) time scale governing the behavior of confined water, which is characterized by two distinct Arrhenius regimes, in full analogy with the present findings for water confined in microgels. This behavior was interpreted as a strong-strong transition due to the coupling of hydration water with the fluctuations of the protein structure and found to occur in correspondence to the protein dynamical transition. The same correspondence is found in our simulations, suggesting that hydration water, through hydrogen bonding, plays a driving role in the dynamical transition. In support of this idea comes the fact that the behavior with concentration is similar for both PNIPAM microgels and hydrated proteins.

\subsection{Comparison between MD simulations and EINS data}
\begin{figure}[htb]
\centering
\includegraphics[width=0.95\linewidth]{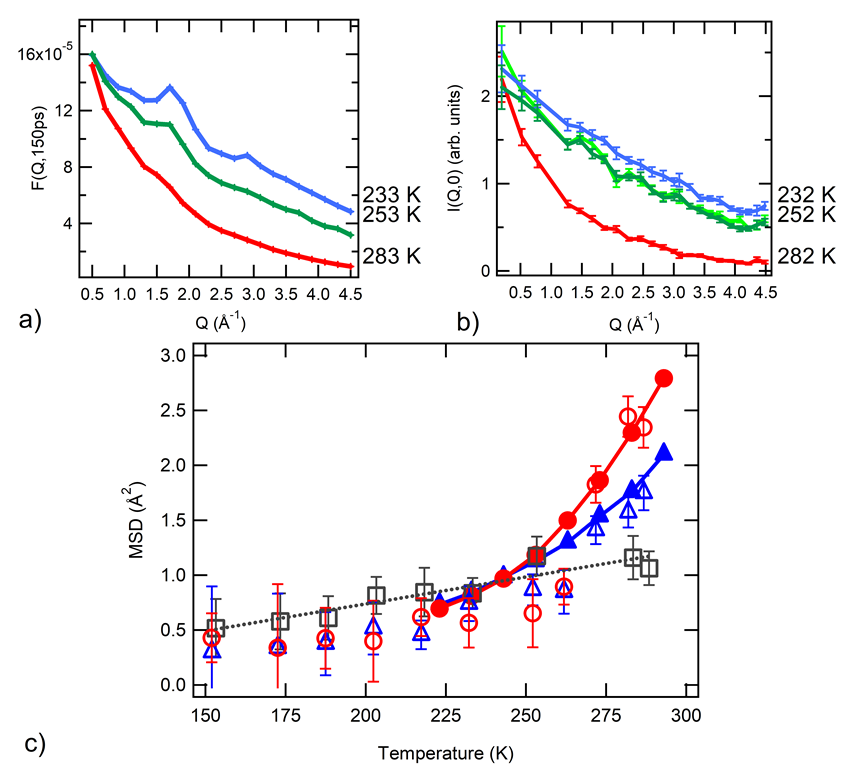}
\caption{\textbf{Quantitative comparison between experiments and simulations.} (a) Neutron spectra as a function of temperature as calculated from the MD simulations of the PNIPAM microgel at 40\%. The total intermediate scattering functions are displayed at a value of 150~ps for comparison with the experimental data. (b) EINS spectra measured on the sample with a PNIPAM concentration of 43\% as a function of  temperature. Data shown at 252~K compare the spectra measured under cooling (dark green) and heating (green). (c)Temperature dependence of experimental (open symbols) MSD for PNIPAM mass fractions 43\% (circles), 60\% (triangles) and 95\% (squares) and numerical (filled symbols) MSD, calculated at 150~ps for PNIPAM hydrogen atoms only, for PNIPAM mass fractions 40\% (circles) and 60\% (squares). The dotted line is a guide to the eye suggesting a linear behavior for the dry sample, for which the dynamical transition is suppressed.}
\label{fig:Iq}
\end{figure}

We now provide a direct comparison between experimental and numerical data. To this end, we have calculated the elastic intensity in simulations, as explained in the Supplementary Materials. The resulting $I(Q,0)$ as a function of $Q$ are shown in Fig.~\ref{fig:Iq} for simulations (a) and experiments (b) at corresponding temperatures for the 40 wt \% sample. The data are in very good qualitative agreement, also showing the presence of more pronounced peaks at comparable wave vectors. From the simulated trajectories we can also calculate the MSD of the PNIPAM hydrogen atoms and compare them with those resulting from the fit of the experimental $I(Q,0)$. The comparison is shown in Fig.~\ref{fig:Iq}(c). The numerical MSD are in striking quantitative agreement with the experimental ones, without need of any scaling factor.

\section{Discussion and Conclusions}
Here we have reported evidence of the occurrence of a dynamical transition in concentrated PNIPAM microgel suspensions at low temperatures, akin to that observed in proteins. We have combined EINS measurements with atomistic MD simulations, based on a new model specifically developed to mimic the interior of a microgel network. While a detailed numerical representation of a whole microgel is currently unfeasible, our study is capable of resolving the intranetwork dynamics of PNIPAM atoms at the short time scales probed by EINS experiments. Thanks to the use of highly concentrated samples, we have extended the investigation of microgel suspensions to unprecedentedly low temperatures, avoiding water crystallization. Our findings are in agreement with calorimetric data for linear PNIPAM chains, showing that crystallization is absent for polymer mass fractions above $\sim\!50\%$~\cite{afroze2000phase,van2004kinetics}. These results open up the possibility of systematically investigating very high concentration and low temperature microgel samples beyond current practice.

The dual colloidal/polymeric nature of microgels~\cite{lyon2012polymer} bridges the properties of classical hard-sphere colloids and soft polymeric particles, allowing the presence of two distinct glass transitions on the colloidal and the polymeric scale, respectively. Both transitions depend on temperature, given the thermoresponsive character of PNIPAM, which is responsible for the so-called volume phase transition at $T_{VPT}\sim 305$~K, from a swollen state of the microgel particles for $T\lesssim T_{VPT}$ to a collapsed state above it. In the swollen regime, a colloidal glass transition takes place at a PNIPAM mass fraction of $\sim10$ wt \%~\cite{bischofberger2015new}. Although very few studies of microgels at high concentrations exist, we can expect the glass transition on the polymeric scale to be similar to that reported for PNIPAM chains, amounting to about 80 wt \% at ambient temperature~\cite{afroze2000phase,van2004kinetics}. This means that, for the concentrations studied in this work, the microgel samples are macroscopically arrested in a glass-like state. A legitimate question thus concerns the reproducibility of our samples and their stability. A reliable preparation and measurement protocol allowed us to obtain homogeneous samples (see Materials and Methods), ensuring a complete reproducibility of the results and a smooth, progressive variation of the dynamical observables with microgel concentration at the investigated short time and length scales.

The study of microgels in water at low temperatures poses a numerical challenge in appropriately identifying a suitable water model, that is able to realistically describe its peculiar behavior. Very accurate models, which are usually able to describe specific aspects, exist. In particular TIP4P/ICE was devised to accurately reproduce the solid properties of water~\cite{tip4pICE}. We have chosen to work with this model because it reproduces water melting at $\sim\!270$~K, much closer to reality than the $\approx\!250$~K predicted by the mostly adopted TIP4P/2005~\cite{tip4p2005}. Thus, it should also be more realistic in describing supercooled water environments. Our results confirm this hypothesis, providing quantitative agreement with experiments in the investigated $T$-range for the dynamical properties of PNIPAM atoms. It will be interesting to test other models under the same conditions, particularly TIP4P/2005, as well as to extend our studies at higher temperatures.

The similarity of the transition reported here for hydrated PNIPAM microgels with that commonly observed in proteins could be important to better understand the microscopic origin of such a transition. There are still controversial aspects in the literature, particularly concerning the role played by water. Two experimental studies have recently proposed an alternative picture for the dynamical transition, which would support the view that a transition may occur independently from the presence of water. Liu and coworkers~\cite{liu2017dynamical} reported the observation of a dynamical transition also in dry deuterated proteins, mainly due to backbone heavy atoms. While the relevance of the traditional transition in hydrated proteins remains unaltered, the role of this ``dry'' transition is still uncertain. It might provide one of the contributions to flexibility involved in biological processes, although in most proteins full functionality is achieved in the hydrated state only.
In addition, Mamontov and coworkers~\cite{mamontov2016} investigated a simple polymeric system in nonaqueous environment, claiming to observe a dynamical transition similar to the protein one. However, in the latter study, the temperature of the transition significantly depends on the polymer concentration, differently from what was found here and for hydrated protein studies, suggesting that its microscopic nature could be different. On the other hand, the present results highlight the role of water-PNIPAM interplay in the dynamical transition, suggesting that all systems with structural complexity and water interactions similar to proteins should exhibit the same phenomenon. A natural and interesting extension of this work will be to investigate other macromolecular environments with different polymeric architectures, such as PNIPAM linear chains, to further ascertain both the occurrence and the microscopic nature of the dynamical transition.

We found that microgels are extremely efficient in confining water, because the lower limit of sample crystallization is close to 40\% PNIPAM mass fraction, a value that exceeds the amount of water what commonly found in other confining environments. The typical protein concentration where crystallization is avoided is about 70\% in protein mass fraction~\cite{doster1989dynamical,smith1990dynamics,wood2008coincidence,schiro2015translational}. Thus, our results suggest that stable samples with a majority of water can be studied down to very low temperatures, which could be of potential interest for the investigation of liquid-like water behavior in the so-called no man's land region of the phase diagram~\cite{gallo2017supercooled,kim2017,woutersen2018}. The efficient confinement role played by PNIPAM microgels is probably due to their intrinsic network disorder and to their inhomogeneous internal architecture. This further motivates the abovementioned investigations on PNIPAM linear chains to shed light on this aspect.

As a whole, our results strongly suggest that the dynamical transition is a generic feature of water hydrating complex macromolecular suspensions with biological and nonbiological implications. This work also puts forward colloidal PNIPAM microgels as excellent models for a deeper understanding of the functional relevance of the protein dynamical transition.
In proteins, an energy landscape endowed with a large number of conformational states, each containing several tiers of substates, is considered of essential biological importance, as proteins could not function without such a reservoir of entropy~\cite{fenimore2004confsub}. Likewise, the occurrence of a protein-like dynamical transition in PNIPAM-based systems might well be ascribed to a similarly rich amount of conformational substates. Future work will aim to elucidate this point and to provide generality to the present results.

\section*{Materials and Methods}
\paragraph*{Sample preparation}
Microgels were synthesized by precipitation polymerization at $T=343$~K of $N$-isopropylacrylamide (NIPAM) in water (0.136~M) in the presence of $N$,$N'$-methylenebisacrylamide (BIS) (1.82~mM). The reaction was carried out in presence of 7.80~mM sodium dodecylsulfate as surfactant and potassium persulfate (2.44~mM) as radical initiator. The reaction was carried out for 10~hours in nitrogen atmosphere. The obtained colloidal dispersion was purified by exhaustive dialysis against pure water, lyophilized, dispersed in D$_2$O, lyophilized and dispersed again in D$_2$O to a final concentration of 10\%. The obtained microgels were characterized by dynamic light scattering (Zetasizer Nano S, Malvern) and the hydrodynamic diameter was found to be $94\pm3$~nm at 293~K with a size polydispersity of $0.17\pm0.01$~nm.

\paragraph*{EINS experiments and stability of the samples}
EINS experiments were carried out on PNIPAM microgel suspensions with a PNIPAM mass fractions of 43, 50, 60, 70, and 95\% (dry sample). To single out the incoherent signal from PNIPAM hydrogen atoms samples were prepared in D$_2$O. Measurements were performed at the backscattering spectrometer IN13 of the ILL. PNIPAM samples were measured inside flat aluminum cells ($3.0 \times 4.0$~cm$^2$), sealed with an In o-ring. The thickness of the cell was adjusted to achieve a transmission of about 88\% for each sample. To obtain samples with the required PNIPAM concentration within the cell used at IN13, we started from the prepared microgel dispersion at 10\%. We then proceeded by evaporation of the exceeding D$_2$O in dry atmosphere using a desiccator under moderate vacuum ($\sim10$~mmHg). Once the final concentration was reached , cells were sealed and left to homogenize at room temperature for at least 4 days before analysis. The sample at 95\% composition was prepared from film casting the PNIPAM dispersion at 10\% up to dryness in petri dish. The obtained transparent films were milled with an IKA MF 10.1. Cutting-grinding gave rise to a rough powder that was poured into an aluminum cell for neutron scattering. The weight of each sample was checked before and after the measurement without observing any appreciable variation. Moreover, after EINS measurements, the cells were opened and no changes in the sample morphology were detected, showing a homogeneous character and no compartmentalization effects. EINS data were acquired in the fixed-window elastic mode, thus collecting the intensity elastically scattered as a function of $Q$. Data were corrected to take into account incident flux, cell scattering, and self-shielding. The intensity of each sample was normalized with respect to a vanadium standard to account for the detector efficiency. Multiple scattering processes were neglected.

\paragraph*{PNIPAM model and MD simulations}
The model mimics, with an atomic detail, a cubic portion of PNIPAM microgel for polymer mass fractions of 40 and 60 wt \%. The macromolecular network, modeled as isotropic, includes 12 atactic PNIPAM chains joined by 6 fourfold bisacrylamide cross-links (see fig.~\ref{fig:model}). Amide groups of PNIPAM residues are represented in the trans conformation. The three-dimensional percolation of the polymer scaffold is accounted for the covalent connectivity between adjacent periodic images. The number average molecular weight of chains between cross-links, $M_c$, is 1584~g/mol, with a polydispersity index of 1.02 and an average degree of polymerization of 14. Taking into account the NIPAM/BIS feed ratio used in the synthesis and the nonuniform cross-link density of PNIPAM microgels, the model represents a region in proximity to the core-shell boundary of the particle. MD simulations of PNIPAM microgels were carried out at 8 temperatures (from 293 to 223~K every 10~K). A trajectory interval of about 0.5~$\mu$s was calculated for each temperature. Additional details are given in the Supplementary Materials.

\section*{Acknowledgments}
We acknowledge ILL for beamtime and CINECA for computer time within the Italian SuperComputing Resource Allocation (ISCRA). \textbf{Founding:} LT, MB, EC and EZ acknowledge support from European Research Council (ERC-CoG-2015, Grant No. 681597 MIMIC), MB, AO and EZ from Italian Ministry of Education, University and Research (MIUR) through a Progetto di Rilevante Interesse Nazionale (PRIN) grant. \textbf{Author contributions:} AO and EZ supervised the research. EB and MB prepared the samples. MZ, MB, EZ performed neutron scattering experiments with FN. EC designed the network model. LT, EC and EZ performed molecular dynamics simulations. MZ and LT analyzed data. MZ, LT, AO and EZ wrote the paper with inputs and suggestions from all authors. MZ, LT, and EB contributed equally to this work. \textbf{Competing interests:} No conflict of interest to declare. \textbf{Data and materials availability:} All data needed to evaluate the conclusions in the paper are present in the paper and/or the Supplementary Materials. Additional data available from authors upon request.

\newpage
\renewcommand{\thefigure}{S\arabic{figure}}
\renewcommand{\thetable}{S\arabic{table}}
\setcounter{figure}{0}

\section*{Supplementary Material for: Evidence of a low temperature dynamical transition in concentrated microgels}

\paragraph*{EINS measurement and data analysis}
The thermal protocol followed during the EINS measurements is reported in Tab.~\ref{thermal}.
\begin{table}[tbhp!]
\centering
	\begin{tabular}{r|cc|ccccc}
        \hline
		Step	&	$T_{i}$	&	$T_{f}$ & $43\%$	& $50\%$	&	$60\%$	&	$70\%$	& $95\%$	\\
					& (K)			& (K)			& (K/min)		& (K/min)		& (K/min)		& (K/min)		& (K/min)		\\
		\hline
		1 		& 288 		& 283 		&	0.7				& 0.3				& 0.7				& 0.6				& 0.3				\\
		2 		& 283 		& 253 		& 0.8				&	0.6				&	0.8				&	0.8				&	0.6				\\
		3 		& 253 		& 233 		& 0.7				& 0.5				& 0.7				&	0.7				&	0.5				\\
		4 		& 233 		& 218 		& 0.6				& 0.4				& 0.6				&	0.6				&	0.4				\\
		5 		& 218 		& 203 		& 0.6				& 0.4				& 0.6				&	0.6				&	0.4				\\
		6 		& 203 		& 188 		& 0.6				& 0.4				& 0.6				&	0.5				&	0.4				\\
		7 		& 188 		& 173 		& 0.5				& 0.4				& 0.6				&	0.5				&	0.4				\\
		8 		& 173 		& 153 		& 0.7				& 0.5				& 0.7				&	0.7				&	0.5				\\
		\hline
		9 		& 153 		& 203 		& 1.2				& 1.7				& 1.2				&	1.2				&						\\
		10 		& 203 		& 253 		& 1.2				& 1.7				& 1.1				&	1.1				&						\\
		11		&	253			&	263			& 0.4				& 0.7				&	0.4				&	0.4				&						\\		
		12		& 263 		& 273 		& 0.4				& 0.7				& 0.4				&	0.4				&						\\
		13 		& 273 		& 283 		& 0.4				& 0.7				& 0.4				&	0.4				&						\\
		\hline
	\end{tabular}
	\caption{\textbf{Temperature protocol followed during the EINS experiment.} Cooling (top part) and heating (bottom part) ratios used for each PNIPAM sample. $T_i$ and $T_f$ represent the initial and final temperature of each step. Samples are labeled with their PNIPAM mass fraction.}
	\label{thermal}
\end{table}

\begin{figure}[h!]
\centering
\includegraphics[width=.75\linewidth]{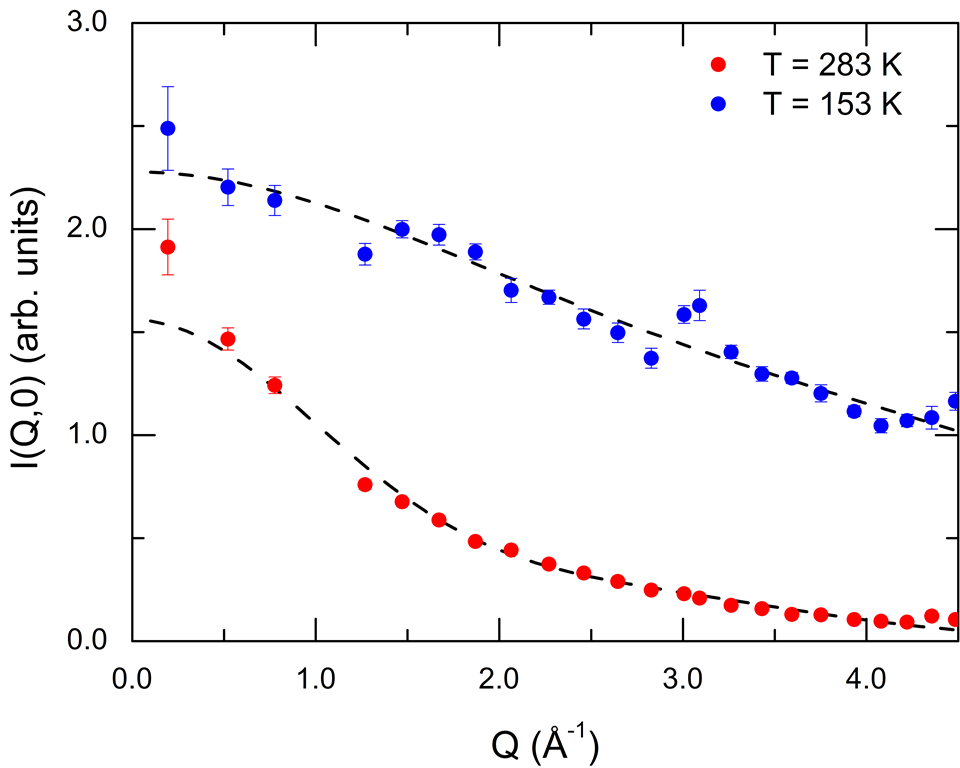}
\caption{\textbf{EINS data for PNIPAM 43 wt \% sample.} $I(Q,0)$ measured on the PNIPAM 43 wt \% sample at 283~K (red dots) and 153~K (blue dots); the black dashed line is the fit using Eq. \protect\ref{Eq:doublewell}.}
\label{fig:Iq40}
\end{figure}

\begin{figure}[h!]
\centering
\includegraphics[width=.75\linewidth]{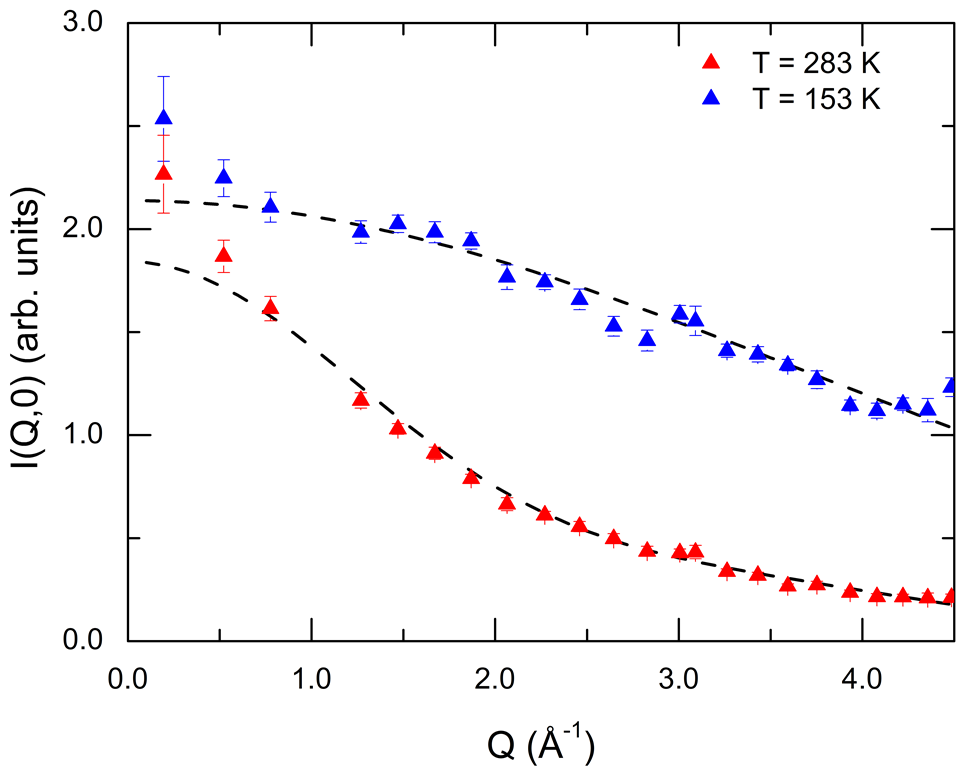}
\caption{\textbf{EINS data for PNIPAM 60 wt \% sample.} $I(Q,0)$ measured on the PNIPAM 60 wt \% sample at 283~K (red dots) and 153~K (blue dots); the black dashed line is the fit using Eq. \protect\ref{Eq:doublewell}.}
\label{fig:Iq60}
\end{figure}

\begin{figure}[h!]
\centering
\includegraphics[width=.75\linewidth]{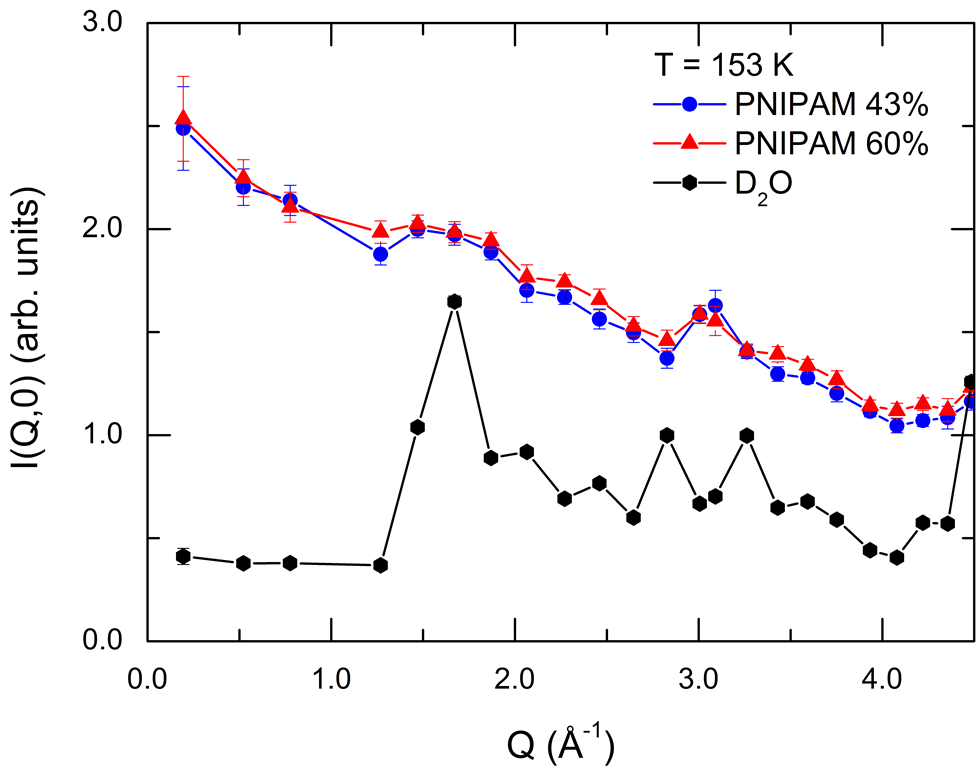}
\caption{\textbf{D$_2$O contribution to EINS spectra.} $I(Q,0)$ measured on pure D$_2$O compared with that obtained on PNIPAM samples with a mass fraction of 43 and 60\% at 153~K.}
\label{fig:D2O}
\end{figure}

In the incoherent approximation, the elastic neutron scattering intensity $I(Q,0)$ can be described by the double-well model \cite{doster1989dynamical}. Within this approximation, hydrogen atoms are supposed to be dynamically equivalent and may jump between two distinct sites of different free energy. The elastic intensity can be thus written as:
\begin{eqnarray}
I(Q,0)&=&\exp\left(-Q^2 \left\langle\Delta u^2\right\rangle_{vib}\right)\nonumber\\
			&&\left[1-2p_1p_2\left(1-\frac{\sin(Qd)}{Qd}\right)\right]
\label{Eq:doublewell}
\end{eqnarray}
where $p_1$ and $p_2$ are the probabilities of finding the hydrogen atom, respectively, in the ground and excited state, $\left\langle\Delta u^2\right\rangle_{vib}$ corresponds to the vibrational mean square displacement of protons rattling in the bottom of the wells, and $d$ is the distance between the two wells.
In this model, where a transition between the two states represents a jump between conformational substates in the free energy surface, the amplitude of the 3-dimensional mean squared displacement (MSD) is given by the relationship \cite{paciaroni2002effect}:
\begin{eqnarray}
\mbox{MSD}&=&-6\left(\frac{d\ln I(Q)}{dQ^2}\right)_{Q=0}\nonumber\\
					&=&6\left\langle\Delta x^2\right\rangle_{vib}+2p_1p_2 d^2
\label{Eq:MSD}
\end{eqnarray}
Typical examples of the fits to the data with Eq. \ref{Eq:doublewell} are shown in Figs.~\ref{fig:Iq40} and \ref{fig:Iq60} respectively for 43 and 60 wt \% samples.
A pure D$_2$O sample was also measured to completely rule out the possibility of its crystallization in PNIPAM samples. Fig.~\ref{fig:D2O} shows the $I(Q,0)$ for heavy water and for PNIPAM 43 and 60 wt \% at 153~K. PNIPAM samples do not show any traces of the D$_2$O Bragg peaks.

\begin{figure}[h!]
\centering
\includegraphics[width=.98\linewidth]{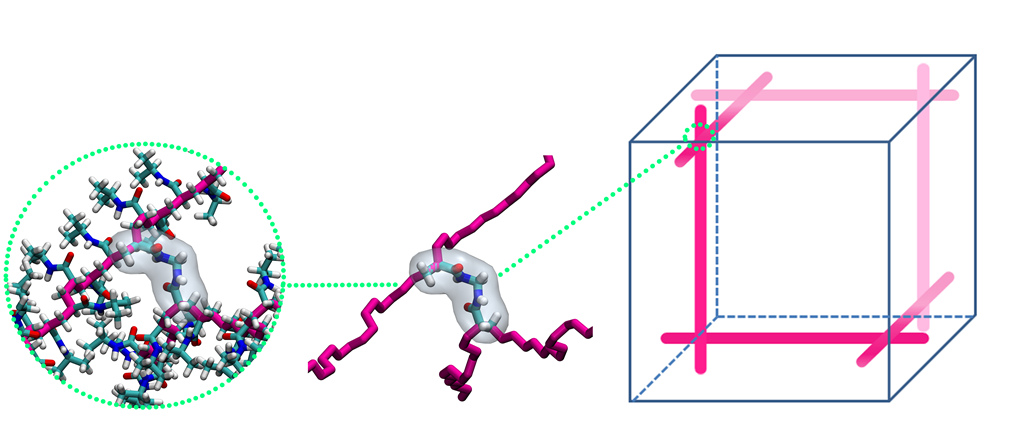}
\caption{\textbf{Schematic representation of the microgel network model.}}
\label{fig:model}
\end{figure}

\paragraph*{PNIPAM model development}
An isotropic polymer scaffold, with the network topology shown in Fig.~\ref{fig:model}, was built by cross-linking atactic PNIPAM chains in the minimum energy conformation \cite{flory1966random}. Amide groups of PNIPAM residues were modeled in the trans conformation. Extra-boundaries covalent connectivity between polymer chains was applied. The network model includes 6 4-fold bisacrylamide junctions and has a number average molecular weight of chains between cross-links, $M_c$, of 1584~g/mol, with a polydispersity index of 1.02. The average degree of polymerization of chains between junctions is $14\pm2$. The polymer scaffold was hydrated by a shell of water molecules to set the PNIPAM concentration in the microgel. Then the system was equilibrated at 293~K in a pressure bath at 1 bar up to a constant density value, i.e. tot-drift less than $2 \times 10^{-3}$~g~cm$^{-3}$ over 20~ns. A similar equilibration procedure was applied at each temperature before the NVT run.

\paragraph*{MD simulations procedure}

Molecular dynamics simulations of PNIPAM microgels were carried out using the GROMACS 5.0.4 software \cite{Pall2015}. The polymer network was modeled using the OPLS-AA force field \cite{Jorgensen1996} with the implementation by Siu et al. \cite{Siu2012}, while water was described with the TIP4P/ICE model \cite{tip4pICE}. The system was equilibrated for 120~ns for $T\geq273$~K and for 320~ns for $T\leq263$~K in the NPT ensemble, taking into account the longer equilibration time needed at lower temperatures.  Simulation data were collected for 330~ns in the NVT ensemble, with a sampling of 0.2~frame/ps.  The leapfrog integration algorithm was employed with a time step of 0.2~fs, cubic periodic boundary conditions, and minimum image convention. The length of bonds involving hydrogen atoms was kept fixed with the LINCS algorithm. The temperature was controlled with the velocity rescaling thermostat coupling algorithm with a time constant of 0.1~ps. Electrostatic interactions were treated with the smooth particle-mesh Ewald method with a cutoff of non-bonded interactions of 1~nm. The last 100~ns of trajectory were considered for analysis. The software MDANSE \cite{MDANSE}
as well as in-house codes were used for analysis of MD simulations to be compared with the neutron scattering data. Trajectory format manipulations (or conversions) were carried out by the software WORDOM \cite{wordom}.
The software VMD \cite{VMD}
was employed for graphical visualization.

\begin{figure}[h!]
\centering
\includegraphics[width=0.98\linewidth]{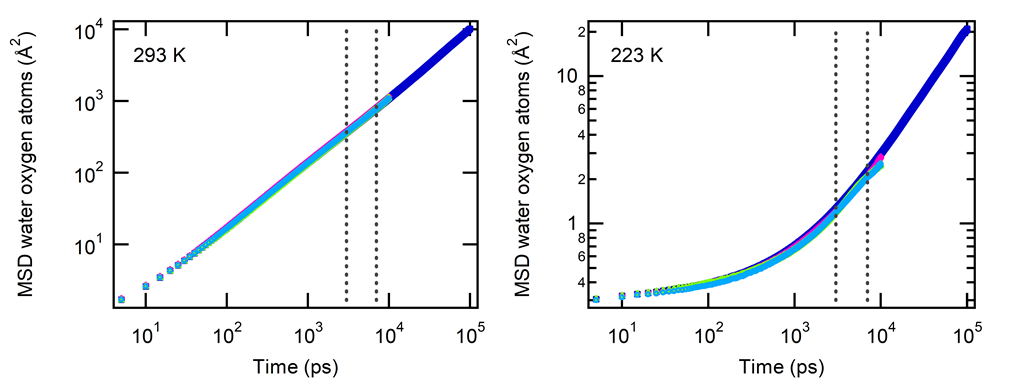}
\caption{\textbf{Aging effect on water dynamics.} Comparison between the mean square displacements of water oxygen atoms at 293~K (left panel) and 223~K (right panel) as calculated from the MD simulations over 100~ns of trajectory (blue squares), 0-10~ns (pink diamonds), 50-60~ns (green triangles), and 90-100~ns (light blue circles). Dashed lines highlight the linear region used for water diffusion coefficient calculation.}
\label{fig:msd_water}
\end{figure}

\begin{figure}[h!]
\centering
\includegraphics[width=0.94\linewidth]{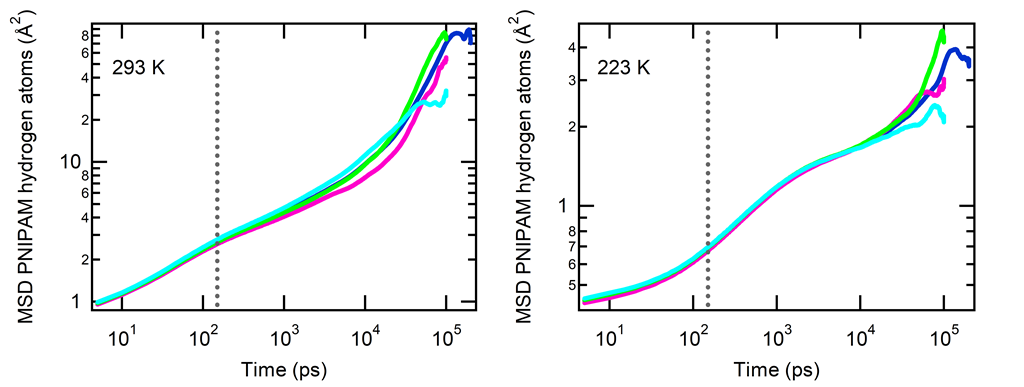}
\caption{\textbf{Aging effect on PNIPAM dynamics.} Comparison between the mean square displacements of PNIPAM hydrogen atoms at 293~K (left panel) and 223~K (right panel) as calculated from the MD simulations over 200~ns of trajectory (blue lines), 0-100~ns (pink lines), 50-150~ns (green lines), and 100-200~ns (light blue lines). The value at the experimental time resolution of 150~ps is indicated with a dashed line.}
\label{fig:msdPNIPAM}
\end{figure}

\begin{figure*}[htb!]
\centering
\includegraphics[width=0.65\textwidth]{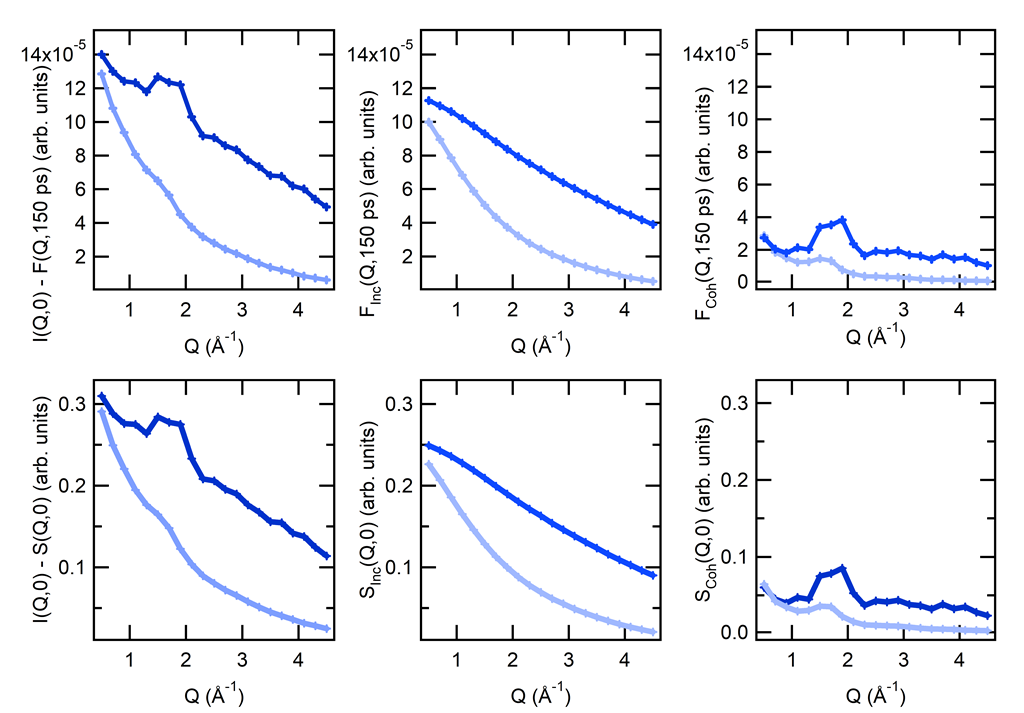}
\caption{\textbf{I(Q,0) from numerical simulations.} Comparison between simulated neutron spectra $I(Q,0)$ calculated from the intermediate scattering functions at the experimental time resolution of 150~ps (upper panels) and from the total dynamical structure factors convoluted with the experimental resolution (lower panels) at 223~K (blue lines) and 293~K (light blue lines).The incoherent $F_{Inc}$ and $S_{Inc}$ are reported in the central panels of each row, while the coherent $F_{Coh}$ and $S_{Coh}$ contributions are shown in the rightmost panels.}
\label{fig:fqtsw}
\end{figure*}

\paragraph*{Reproducibility of numerical results}
The numerical data are not affected by aging dynamics on the timescales investigated in the manuscript. To show that this is the case we report the evolution of the MSD upon changing the time interval for the calculation  as well as waiting time $t_w$. Fig.~\ref{fig:msd_water} shows the MSD of water oxygen atoms  at the highest ($T=293$K) and lowest ($T=223$~K) investigated temperatures. We calculate the MSD for a total time of 100~ns and compare it to the corresponding one calculated for only 10~ns (as used in the manuscript, Fig.~\ref{fig:pnipam}(c). In addition, we also calculate it for $t_w=0,50,90$~ns. All curves superimpose on the entire investigated timescale, in particular for the intermediate window where the self-diffusion coefficient was extracted, indicated by vertical dashed lines. This behaviour holds for both high and low temperatures.
Similarly, Fig.~\ref{fig:msdPNIPAM} shows the MSD of PNIPAM hydrogen atoms at the same two temperatures over a total interval of 100 and 200~ns, the former being the ones from which we extract the value at 150~ps, reported in the manuscript in Fig.~\ref{fig:Iq}(c). We also calculate it for waiting times $t_w=0,50,100$ns.  At long times, some differences, that are mostly attributable to statistical error rather than aging, are visible. However, it is clear that for the timescale of relevance for comparison with experimental data, i.e. 150~ps, indicated by a vertical dashed line, no significant aging effects are found.
This analysis ensures that the numerical data are reproducible and the system is well equilibrated for the short  timescales studied in this work.

\paragraph*{Calculation of $I(Q,0)$ from numerical simulations}
We have calculated both the coherent and incoherent intermediate scattering functions for all atoms in the simulations for wavevectors in the range $ 0.3 \leq Q\leq 4.5$~\AA$^{-1}$. By summing the two contributions, weighted by the appropriate scattering lengths and by the partial concentrations, we have obtained the full differential cross-section. Its value at the experimental time resolution of 150~ps provides the numerical $I(Q,0)$. We have further checked that this procedure yields identical results as the calculation of the total dynamical structure factor (in frequency space) convoluted with the experimental resolution. An example of this procedure is provided in Fig.~\ref{fig:fqtsw}.

\end{document}